\documentclass[12pt,preprint]{aastex}
\font\apj=cmcsc10

\def\ms{PSR~J0437--4715}
\def\sey{RX~J0437.4--4711}
\def\asca{{\it ASCA\/}}
\def\euve{{\it EUVE\/}}
\def\ro{{\it ROSAT\/}}

\def\ls{\lower 2pt \hbox{$\;\scriptscriptstyle \buildrel<\over\sim\;$}} 
\def\gs{\lower 2pt \hbox{$\;\scriptscriptstyle \buildrel>\over\sim\;$}} 
\def\gsim{\lower 2pt \hbox{$\, \buildrel {\scriptstyle >}\over {\scriptstyle
\sim}\,$}}
\def\lsim{\lower 2pt \hbox{$\, \buildrel {\scriptstyle <}\over {\scriptstyle
\sim}\,$}}
\slugcomment{}

\shorttitle{EUVE Light Curves of Seyfert Galaxies}
\shortauthors{J. P. Halpern et al.}

\begin{document}

\title{An Extreme Ultraviolet Explorer Atlas of Seyfert Galaxy Light Curves: 
Search for Periodicity}

\author{J. P. Halpern}
\affil{Department of Astronomy, Columbia University, New York, NY, 10025-6601}

\author{K. M. Leighly}
\affil{Department of Physics and Astronomy, The University of Oklahoma,
440 W. Brooks St., Norman, OK 73019}

\author{H. L. Marshall} 
\affil{Center for Space Research, Massachusetts Institute of Technology, 77 Massachusetts Avenue, NE80, Cambridge, MA 02139-4307}

\begin{abstract}
The Deep Survey instrument on the {\it Extreme Ultraviolet Explorer}
satellite (\euve) obtained long, nearly continuous soft X-ray light curves of
5--33 days duration for 14 Seyfert galaxies and QSOs.
We present a uniform reduction of these
data, which account for a total of 231 days of observation.
Several of these light curves are well suited to a search for
periodicity or QPOs in the range of hours to days that might be expected
from dynamical processes in the inner accretion disk around
$\sim 10^8\,M_{\odot}$ black holes.  Light curves and periodograms of the
three longest observations show features that could be transient periods:
0.89 days in \sey, 2.08 days in Ton~S180, and 5.8 days in 1H~0419--577.  
The statistical significance of these signals is estimated
using the method of \cite{tk95}, which carefully
takes into account the red-noise properties of Seyfert light curves.
The result is that the signals in \sey\ and Ton S180 exceed 95\% confidence
with respect to red noise, while 1H~0419--577 is only 64\% significant.
These period values appear unrelated to the length of the observation, which
is similar in the three cases, but they do scale roughly as the
luminosity of the object, which would be expected in a dynamical
scenario if luminosity scales with black hole mass.

\end{abstract}


\section{Introduction}

Some Seyfert galaxies have shown indications of transient or quasi-periodic
oscillations with periods of order 1 day.  The first such
evidence was obtained in a 20 day \euve\ observation of the
Seyfert galaxy \sey\ \citep{hm96}, where a possible
signal at a period of $0.906 \pm 0.018$~days was detected.
If attributed to
relativistic beaming or other projection effects of orbiting structures
at a radius of $6\,GM/c^2$, a 0.9~day period
requires a black hole mass of $1.7 \times 10^8\,M_{\odot}$.
Since this is the time scale on which one might expect
to find periods or QPOs corresponding to orbital motion or other
dynamical processes
in the inner accretion disk around supermassive black holes in AGNs,
it is reasonable to hypothesize that such signals may be common,
and that the dearth of continuous X-ray observations of sufficient duration
has prevented them from being recognized until now.

X-ray observations of AGNs that would be long enough to
convincingly detect such periodic signals are rare.
Although several long light curves of quasars
near the ecliptic poles were obtained by the \ro\ All-Sky Survey,
those objects
did not vary during the monitoring period \citep{khs93,t95}.
The {\it EXOSAT\/} long-look observations of Seyfert galaxies (e.g.,
Green, McHardy, \& Lehto 1993) provided continuous light curves, 
but the longest was 3 days in duration so it did not adequately sample
variability on time scales of $\sim 1$~day.  There have
been several analyses of possible QPO features with periods of 1~hr
or less in the {\it EXOSAT\/} observations of NGC~4051 and NGC~5548
\citep{bo94,pl93,pl95},
although the NGC~5548 claim has not withstood critical analysis
\citep{tbi96}. \cite{en98} did not detect any periodicity in a
continuous four-day observation of NGC~3516 with {\it RXTE\/}.
The many ``reverberation-mapping''
campaigns have been longer in duration, but most were not sensitive
to X-ray variability on time scales of hours to a few days.
One exception is the
month-long {\it RXTE\/} observation of NGC~7469 \citep{n98},
which did not, however, reveal any short-term periodicity.
A possible indication of periodicity was obtained in a five-day
\asca\ observation of the Seyfert galaxy {\it IRAS\/}~18325--5926.
A signal at 16~hr was seen in those data \citep{i98}.
\cite{lfr00} pointed to a possible 33~hr
period in an 8 day observation of MCG--6--30--15 with {\it RXTE\/}.
\cite{bkt01} claimed that an 8~hr observation of Mrk~766 by
{\it XMM-Newton} shows a period of 4200~s, although its significance
according to the analysis of \cite{bwe01} is not very high.

\section {EUVE Observations of Seyfert Galaxies}

The \euve\ Deep Survey/Spectrometer (DS/S) telescope with
Lexan/Boron filter was used
for most of the pointed observations in the Guest Observer
phase of the mission, from 1993 January through 2001 January.
A description of the instruments on \euve\ and their performance
can be found in \cite{svf97}.
Although the DS Lexan band is sensitive in the range 67--178~\AA ,
photons are detected from extragalactic sources
only at the short wavelength
end because of the steep increase in interstellar absorption
as a function of wavelength.  For example \cite{hmm96}
show the effective energy distribution of detected photons in the DS 
for a range of power-law source spectra.
Since nearly all of the detected flux is in the range 70--100 \AA\
(0.12--0.18 keV), we refer to this band,
according to convention, as soft X-rays.

It may not be well known that the \euve\ Deep Survey imager (DS)
made more long X-ray observations of Seyfert galaxies
than any other X-ray satellite.  \euve\ made 23 nearly continuous
observations, interrupted only by Earth occultation, of 14 Seyfert galaxies
and QSOs.  The duration of these pointings
ranged from 3 to 33 days.  In total, they
account for some 231 days of elapsed time.  A log of the
observations presented here is given in Table \ref{tbl1}.
(We do not include in this paper
many shorter \euve\ observations of additional Seyfert galaxies that
were detected in the DS.)
These do not represent a complete or unbiased
sample of Seyfert galaxies in any sense.  Most were chosen mainly
on the basis of their detectability with \euve, a rare
prospect whose chances are improved by selecting targets having 
small Galactic and intrinsic absorbing column densities.  Also, famous
Narrow-line Seyfert 1 Galaxies (NLS1s) are heavily represented
(NGC 4051, Mrk 478, Ton~S180, RE J1034+396)
because they are bright, soft X-ray sources with steep
X-ray spectra \citep{l99b}.  

\begin{figure}
\vskip -0.25 truein
\center{\scalebox{1.00}{\plotone{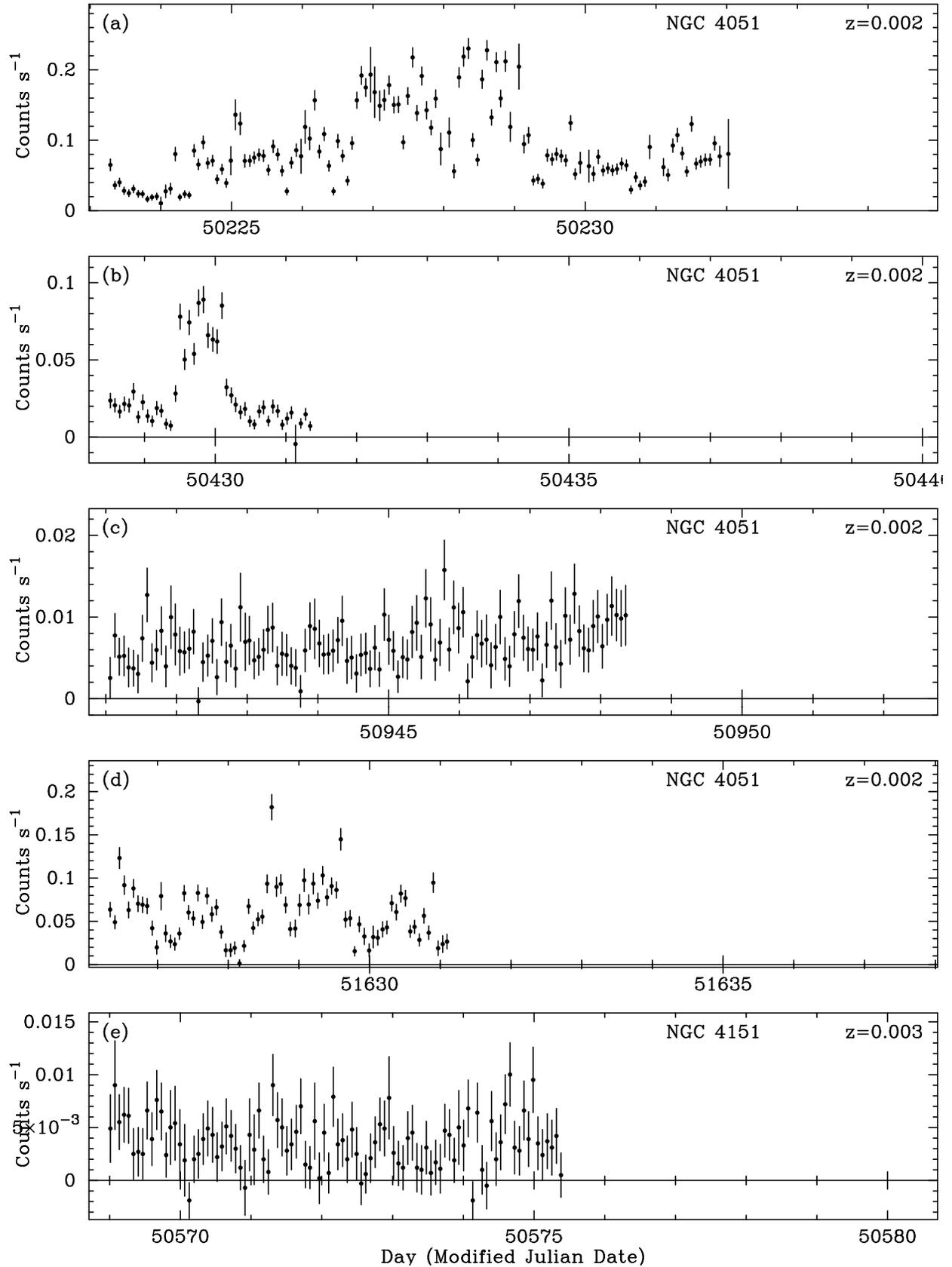}}}
\vskip -0.2 truein
\caption{\euve\ DS light curves of Seyfert galaxies.
Each point represents one satellite orbit.  Background has
been subtracted, and count rates
are corrected for dead-time and telemetry sharing.\label{lc1}}
\end{figure}

\begin{figure}
\vskip -0.25 truein
\center{\scalebox{1.00}{\plotone{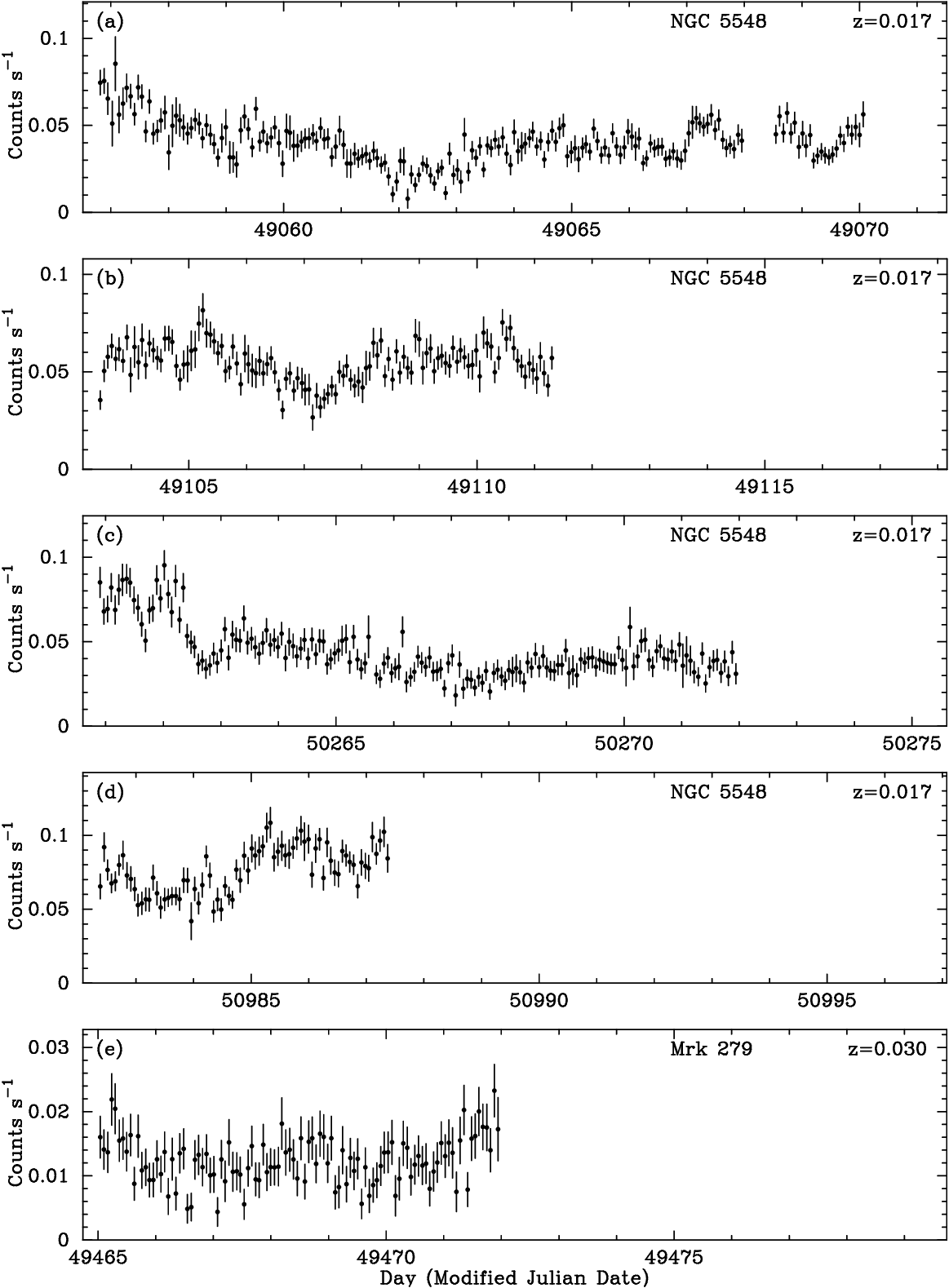}}}
\vskip -0.2 truein
\caption{Same as Figure \ref{lc1}.\label{lc2}}
\end{figure}

\begin{figure}
\vskip -0.25 truein
\center{\scalebox{1.00}{\plotone{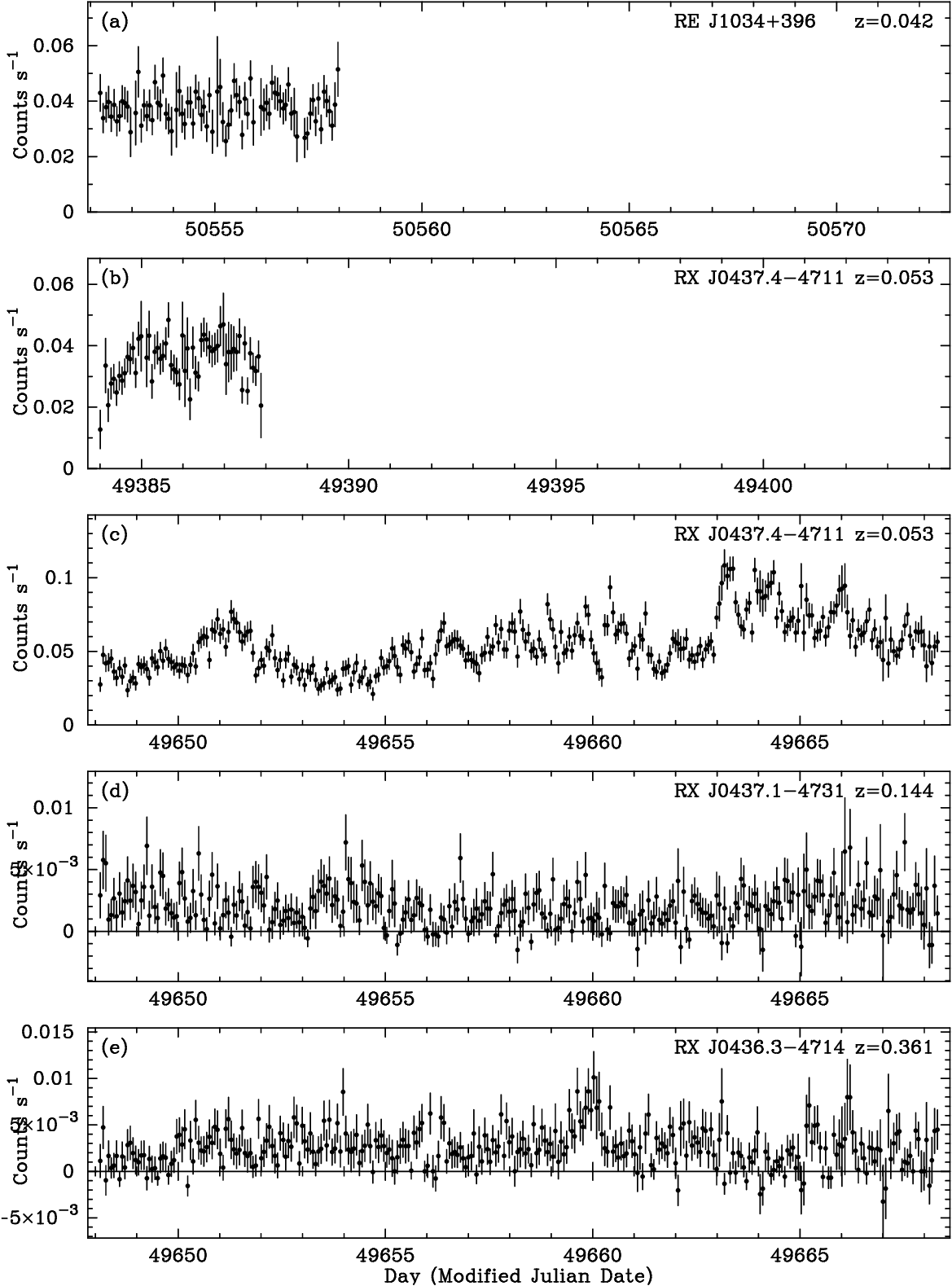}}}
\vskip -0.2 truein
\caption{Same as Figure \ref{lc1}.\label{lc3}}
\end{figure}

\begin{figure}
\vskip -0.25 truein
\center{\scalebox{1.00}{\plotone{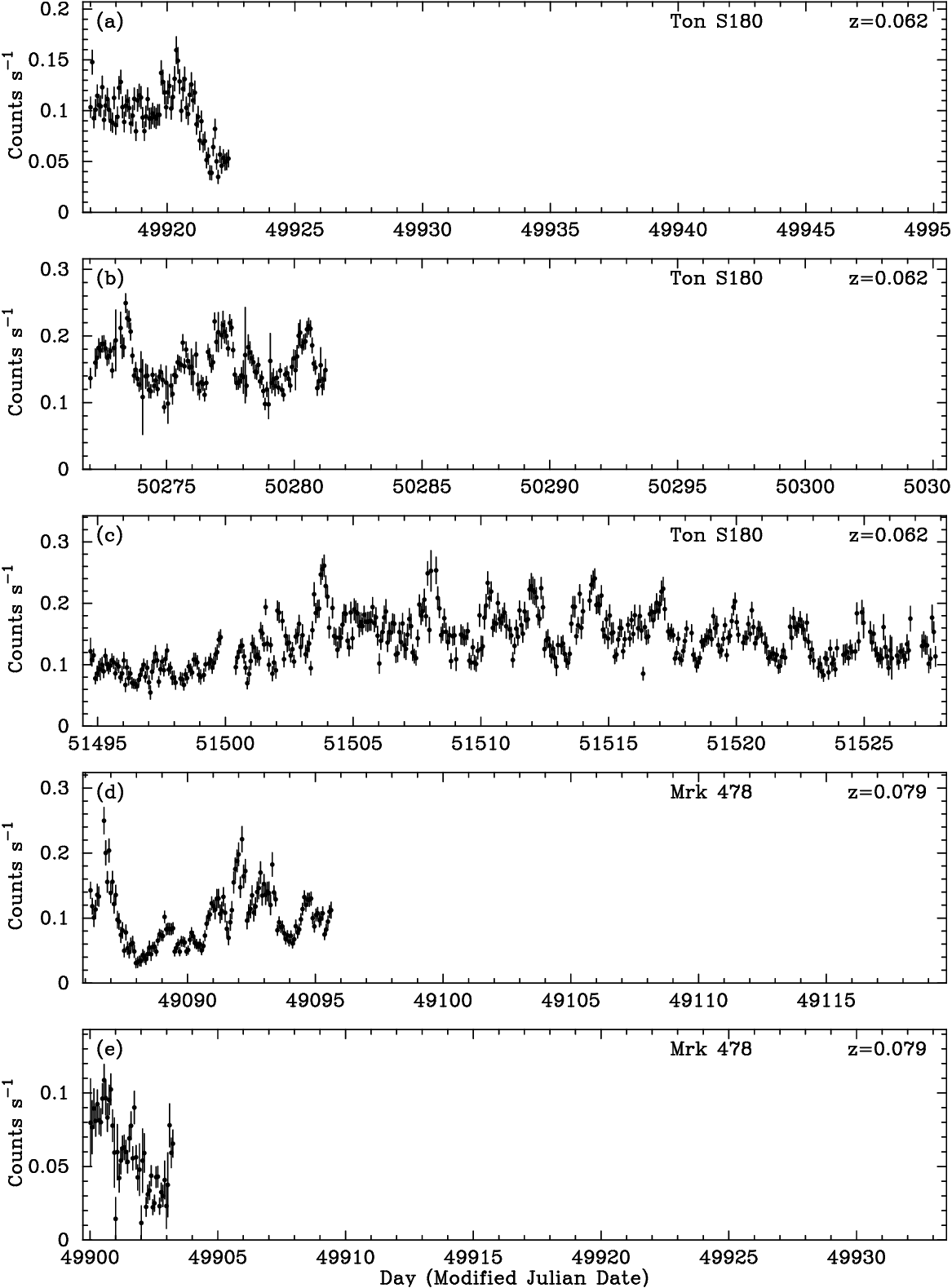}}}
\vskip -0.2 truein
\caption{Same as Figure \ref{lc1}.\label{lc4}}
\end{figure}

\begin{figure}
\center{\scalebox{1.00}{\plotone{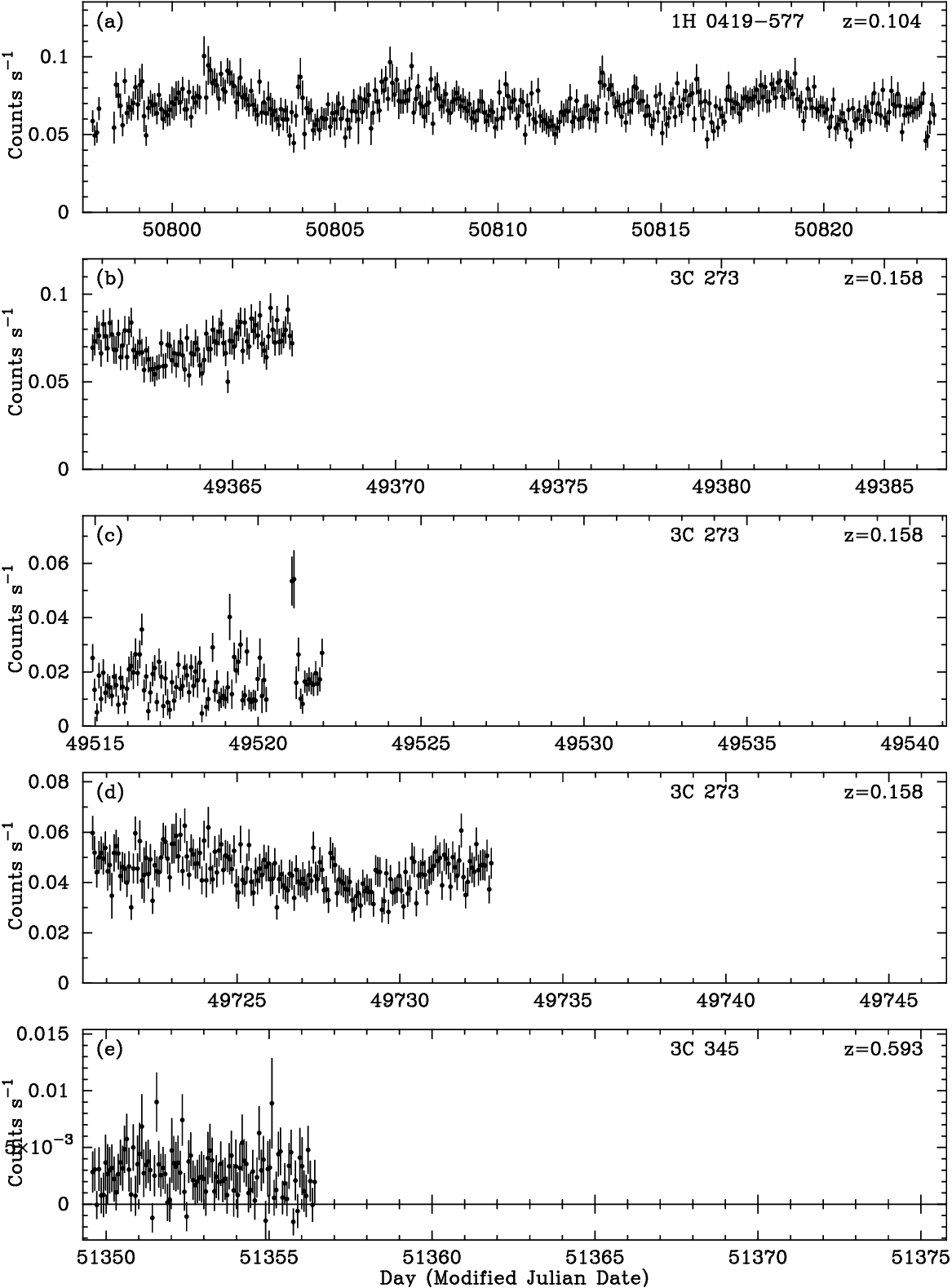}}}
\caption{Same as Figure \ref{lc1}.\label{lc5}}
\end{figure}

\begin{figure}
\center{\scalebox{1.00}{\plotone{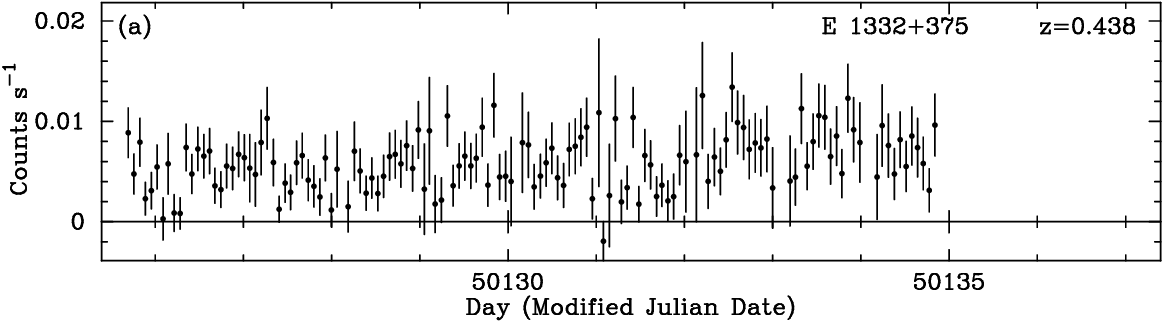}}}
\caption{Same as Figure \ref{lc1}.\label{lc6}}
\end{figure}

Many of these data sets have been published,
at least in part, by the authors listed in the notes to
Table \ref{tbl1}.  Few of these papers,
however, were concerned with power-spectrum analysis or
evidence for periodicity.  Several papers presented spectra of the Seyferts
from the \euve\ short-wavelength spectrometer,
a heroic effort that, unfortunately, yielded
results that only a mother could love.  However, the
corresponding long DS light curves, which are of a quality ranging
from mediocre to excellent, deserve a separate and comprehensive
analysis.  For this paper, we simply extracted a light curve from the
DS imager using a circle whose radius was chosen to include at least
95\% of the source counts, and a surrounding annulus for background subtraction.
Correction factors for dead-time and Primbsching (variable loss of photons
when the count rate is high because of designed
telemetry sharing between the instruments on \euve) were also applied.
These effects are discussed further in \S3.2.
When such a light curve is binned into one point per satellite
orbit, as in Figures \ref{lc1}--\ref{lc6}, it constitutes a uniformly
sampled time series for which an ordinary periodogram
can be calculated to search for features from 3 hours to
several days.  

\section {Searching for Periodicity}

\subsection {Preliminary Requirements}

It important to explain what is meant here by ``periodicity''.
Since AGN power spectra are usually dominated by red
noise, it is difficult to define a rigorous statistical
test for periodicity, since any such feature is superposed on a continuum
of power that is itself poorly characterized.  Some calculations 
of statistical significance in the AGN literature are grossly wrong,
since they are tests against the hypothesis of white noise, or worse,
a constant source.  Similarly, it is not possible to
establish a period in a Seyfert galaxy by marking off two or three
peaks in a light curve and declaring their separation to be ``the period''.
Accretion-powered sources flicker; the dominant feature of flickering
is often an apparent
variation of two or three ``cycles'' over the span of the observation.

We employ some simple criteria to identify plausible candidate periods.
Most important,
we require that there be a narrow peak or a QPO in a periodogram that
1) is clearly separated from the low-frequency noise, and
2) stands out high above the surrounding points.
Some of the candidates to be reported here meet these criteria for
significance, but they are not secure enough if only because the
signal is not clearly present throughout
the entire observation.
However, theory accommodates or even favors
{\it transient} periods in Seyfert galaxies,
since orbiting hot spots or other asymmetric 
structures in an accretion disk {\it should\/} have
finite lifetimes due to Keplerian shear.  Such quasi-periods
could, in principle, be confirmed by observing repeated occurrences.
A third requirement is that the number of cycles detected (coherence time)
should be large.  One would like to see
at least 10 cycles of variation to prove the existence of a period;
20 cycles would be quite convincing.
These requirements (many cycles, allowing for transience)
lead to an ideal observing time of 1 month
to search for periodicity of order 1--2 days.
Some of the light curves in Figures \ref{lc1}--\ref{lc6}
show variations of order unity on 1--2 day time scales,
and so are highly suited to such a search.  In \S 4, we
identify candidate periods from the periodograms of three
objects, and in \S 5 we describe the results of the best formal
analysis of statistical significance that is available for such data.

\subsection {Instrumental Effects}

There are no technical problems that prevent the search for
periodicities of order 1~day with \euve.  Most of the targets
were well detected, and in most cases their variability amplitudes
are larger than their counting statistics, which in turn are larger
than background and systematic effects.  The signals to which we
are sensitive would constitute a major feature of the light curve,
not a subtle effect.
We always take special care to apply and evaluate
dead-time and Primbsch corrections.
These losses are usually due to high background
radiation rates that are further enhanced in the vicinity of the
South Atlantic Anomaly (SAA).  We use only data obtained when
the satellite is on the night side of the Earth.
Most notably, the results that
we describe below are not contaminated by a periodic signal
at 0.99 days that can be generated by passages through
the SAA, even while there is a strong
modulation in the dead-time at this period.  The expected values of
such artificially induced periods are well known,
and they are easily recognized
and eliminated in a power-spectrum analysis of a long observation,
as we describe below.  It is fortunate that the intrinsic variability
amplitude of our targets is large and reliably measured, which we have
verified though analyses of additional sources in the fields
of these targets that are either constant, or show uncorrelated
variability \citep{hm96,hmm96,hlm98}.  In summary, we have identified
no instrumental or terrestrial effects that could be responsible for
the candidate periods described below.

\section {Results on Individual Objects}

\subsection{NGC 4051}

NGC 4051 was the first Seyfert galaxy for which variability
on time scales as short as 150~s was seen \citep{mhm83}.
It has the lowest luminosity of the Seyfert galaxies
observed by \euve, and
it is probably not a coincidence that it exhibits the most
rapid and large-amplitude variability, in addition to extended
low states in which the variability is reduced \citep{ump99}. 
Its EUV flux was well correlated with simultaneous X-ray variations
as observed by {\it RXTE\/}, with no measurable time lag.
In the Comptonization model for the hard X-rays, this implies
that the Comptonizing region is smaller than $20\,R_{\rm Sch}$ for
$M_{\rm BH} > 10^6\,M_{\odot}$ \citep{ump00}. NGC 4051 is the
one object that varies too rapidly to be adequately sampled once
per \euve\ orbit, so we did not calculate a periodogram for it.
A periodogram for the 1996 May observation was presented by
\cite{ca99}, but it shows only features at the satellite orbital
frequency and its first harmonic.

\subsection{NGC 4151}

Although NGC 4151 is one of the brightest Seyfert 1 galaxies in
optical and X-rays, it is also highly absorbed at soft X-ray energies,
so it is not surprising that it was a very weak \euve\ source.

\subsection{NGC 5548}

The observation of NGC 5548 in 1998 June was conducted simultaneously 
with \asca\ and {\it RXTE\/} \citep{crb00}.  These authors
noted that, contrary to some naive expectations,
the variations in the DS light curve actually
seem to {\it lead} similar modulation at harder X-ray energies
by 10--30 ks. They suggested that the EUV emission is indicative
of the underlying physical variability of the source,
and that the optical through EUV portion of the spectrum provides
the seed photons for the production of hard X-rays via thermal Comptonization.

\subsection{RX J0437.4--4711}

\sey\ is an ordinary Seyfert 1 galaxy that
was detected serendipitously in a 20 day \euve\ observation
in 1994 November and December
of the millisecond pulsar \ms, which lies only $4^{\prime}$ away.
A shorter observation of the same field was performed in 1994 January.
The small Galactic column in this direction enhances the detectability
of both objects.
Results on the pulsar were published by \cite{hmm96}, and the
Seyfert galaxy by \cite{hm96}, in which a possible period at 0.906~days was
noted.   A periodogram very similar to
Figure \ref{ps1}a, which is based on the final archived data,
appeared in the latter paper.
The structure that is responsible this signal in \sey\ is evident
in its light curve in Figure \ref{lc3},
especially in the second half of the observation.
At least 11
cycles are present during the second half of the observation.
Here, we note that the periodic signal
is enhanced in the periodogram if only the second half of the
observation is included (Figure \ref{ps1}b).
The shift of the location of the peak from 0.906~days to 0.89~days
when only the second half of the observation is used is within the
expected range of uncertainty, $\Delta P \sim 0.2 P^2/T \approx 0.016$ days,
where $T$ is the duration of the observation.

\subsection{RX J0437.1--4731 and RX J0436.3--4714}

These weak sources appear in the same field as \sey\ and \ms.
They are also listed as unidentified,
EUVE J0437-475 and EUVE J0436--472, respectively,
in the Second \euve\ Source Catalog \citep{bll96}.
Armed with precise positions from the \ro\ High Resolution Imager,
we identified both of these sources spectroscopically as NLS1s using
the CTIO 1.5m telescope in 1998 October and 1999 November.  Their
reshifts are $z = 0.144$ for RX J0437.1--4731 and $z = 0.361$ for
RX J0436.4-4714. 

\subsection{Ton~S180}

Ton~S180 is a NLS1 galaxy.
The 33 day \euve\ observation of Ton~S180 in late 1999 
was coordinated with \asca\ and {\it RXTE\/}
during the last 12 days \citep{etp02,tu02}.  The light
curves during the 12 day period of overlap are well correlated,
with no apparent time delays.
\cite{l99a} showed that NLS1s are more variable than ordinary
Seyfert galaxies of similar X-ray luminosity, and the \euve\
light curves of Ton~S180 (and Mrk 478, another NLS1) seem to
bear this out.  The 1999 \euve\ observation of Ton~S180
was the only one in our
compilation to suffer significantly from imperfect correction
of dead-time and Primbsching due to high background.
However, the symptom of this problem is easily recognized and eliminated.
When the background is high and the correction factor is $\gsim 2$,
the correction becomes inaccurate, possibly nonlinear,
and one or two orbits of data per day are
clearly discrepant from the others.  These bad points induce
a signal in the periodogram at 0.99 days (and its harmonics),
which is the period at which the SAA on the rotating Earth
passes through the night side of the slowly precessing satellite orbit.
(Data obtained when the satellite is on the daylight side of the Earth
are not used here).
When the bad points, in this case less than 10\% of
the data, are excised, the 0.99 day signal
in the periodogram disappears.  Figure \ref{lc4} shows
the light curve cleaned of bad points.

The periodogram of the cleaned 1999 observation of Ton~S180 (Figure \ref{ps2})
shows several possible periods, the strongest being at 2.08 days.  
It is important to note that because this observation
was so long, the detected period cannot be attributed to a
sub-harmonic of the 0.99 day SAA period, as it is well resolved
from it.  Furthermore, the 2.08 day signal is much stronger than
any weak residual at 0.99 days ($1.18 \times 10^{-5}$~Hz), the
latter being undetected in the periodogram of the cleaned light curve.
Weaker peaks in Figure \ref{ps2} are present at 2.81 days and 1.33 days,
again, with no known instrumental effect that could be responsible.

\subsection{Mrk 478}

The light curve of the 1993 April observation of Mrk 478 is taken
directly from \cite{mcs96}.  Because part of the observation was
performed with the source on the boresight position and part with
the source off-axis, corrections had to be applied for the difference
in effective area to produce a continuous light curve.  This procedure
was described in \cite{mcs96}.  That paper also showed how the spectral
energy distribution of Mrk 478 peaks in the EUV.

\subsection{1H 0419--577}

The \euve\ observation of 1H 0419--577 was the second longest in
this compilation, 26 days.  Its periodogram,
shown in Figure \ref{ps3}, has a peak at 5.8~days.
Similar to the case of Ton S180, there may be weaker signals
at other frequencies.
Interestingly, 1H 0419--577 does not appear to suffer
as much from the red noise that dominates \sey, Ton~S180, and
other Seyfert galaxies at low frequencies; thus
the feature at 5.8~days is more prominent.
It is not clear
what is responsible for this qualitatively different behavior, or
indeed if it is a persistent property of the power spectrum of the source.
It is possible that the 5.8~day signal simply {\it is} the red-noise
behavior of this object, as the observation only spans $\approx 4.5$ cycles
of that period.  This light curve of 1H 0419--577
was published without any analysis or interpretation in a paper
about the serendipitous discovery of a new AM Her star
only $4^{\prime}$ from the Seyfert galaxy in this
observation \citep{hlm98}.  The latter paper illustrated
that \euve\ can reliably discover periods.  In the case of the AM Her
star, the period of 85.821 minutes was unambiguous despite its proximity
to the satellite orbit period (96.4 minutes) because the observation was so long.
This may be the only \euve\ observation in which {\it two} new periods
were discovered.

\subsection{3C 273}

The variability amplitude of this high-luminosity QSO is small on short time scales.
Note that one observation in 1994 June was badly compromised by placement of the source on
a ``dead spot'' in the DS microchannel plate that had been created by an earlier
observation of the bright EUV source HZ~43.  The reduced count rate and large fluctuation
during this observation (Figure 5c) is evidently due to this unfortunate placement,
and should be disregarded.

\subsection{E 1332+375}

This QSO was discovered \citep{re82} as a serendipitous X-ray source in an {\it Einstein}
observation of the RS CVn type star BH CVn.  It was then detected in an
\euve\ pointing at the same star \citep{ch99}.  The small Galactic column in this direction,
$N_{\rm H} = 9.2 \times 10^{19}$~cm$^{-2}$, undoubtedly contributes to its detectability.

\section {Formal Test for Periodicity}

\begin{figure}
\center{\scalebox{1.0}{\plotone{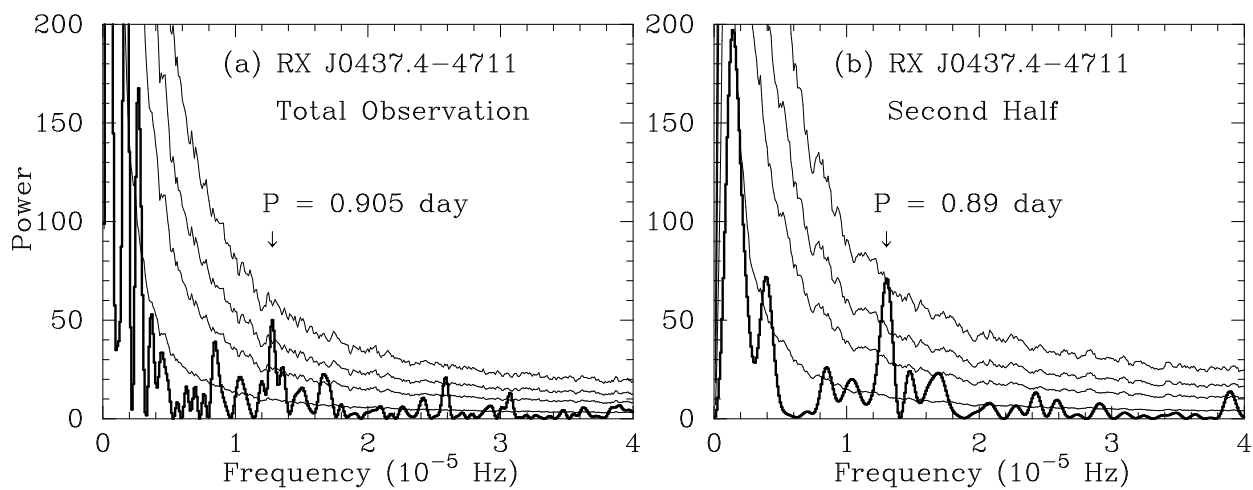}}}
\caption{({\it a}) Periodogram of the entire 20 day DS light curve of \sey ,
from \cite{hm96} ({\it dark line}). A possible signal at 0.906 days is marked.
({\it b}) Periodogram of the second half of the observation, showing 
enhanced significance of the signal, now peaking at 0.89 days.
In both panels, light lines are the significance levels calculated from
1,000 simulated periodograms as described in the text.
From bottom to top, the lines represent 68\%, 95\%, 99\%, and 99.9\%
significance for a signal that exceeds that level of power.
\label{ps1}}
\end{figure}

\begin{figure}
\center{\scalebox{1.0}{\plotone{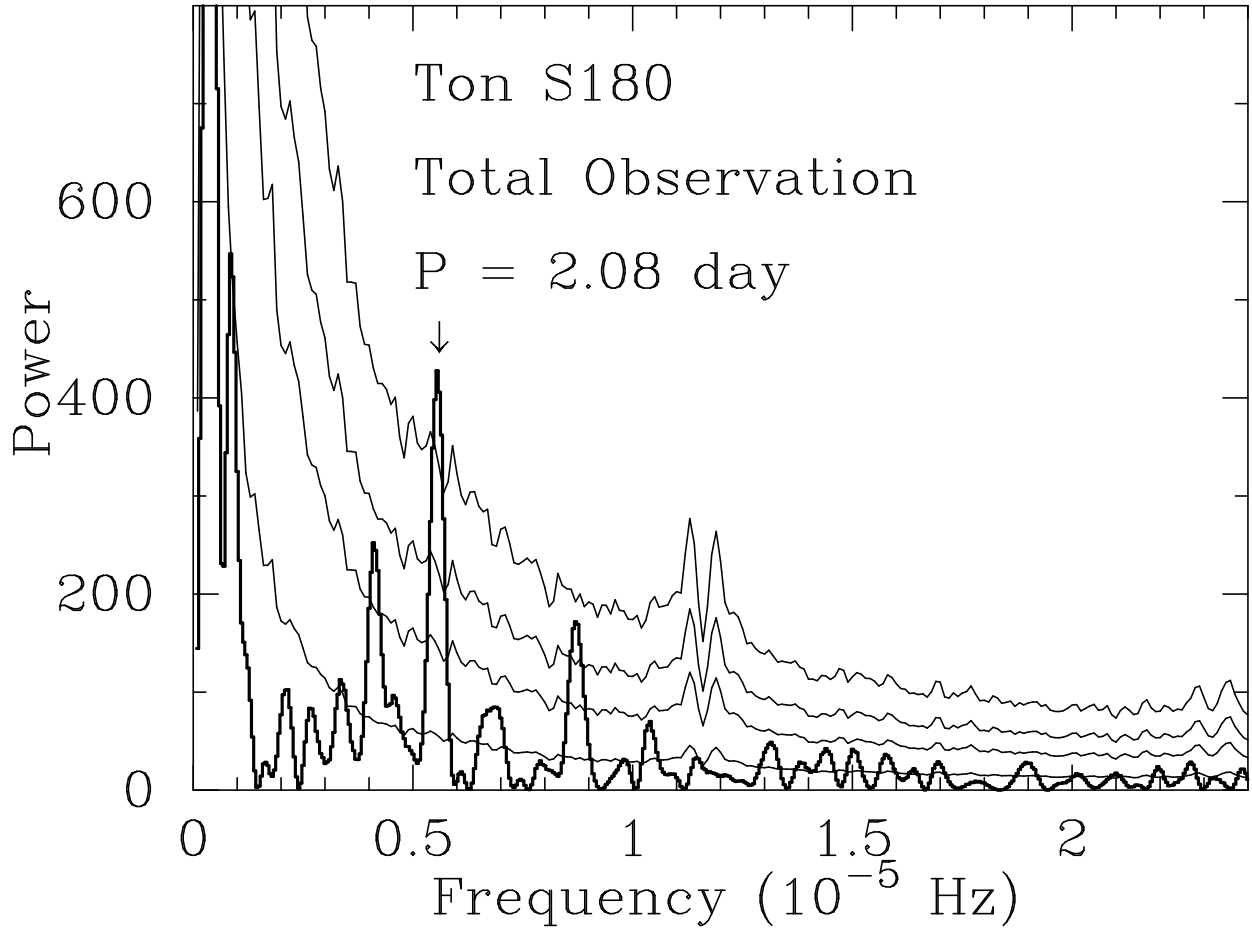}}}
\caption{Periodogram of the long \euve\ DS light curve of
Ton~S180 in 1999 Nov--Dec. Lines have the same meaning as in Figure \ref{ps1}.
\label{ps2}}
\end{figure}

\begin{figure}
\center{\scalebox{1.0}{\plotone{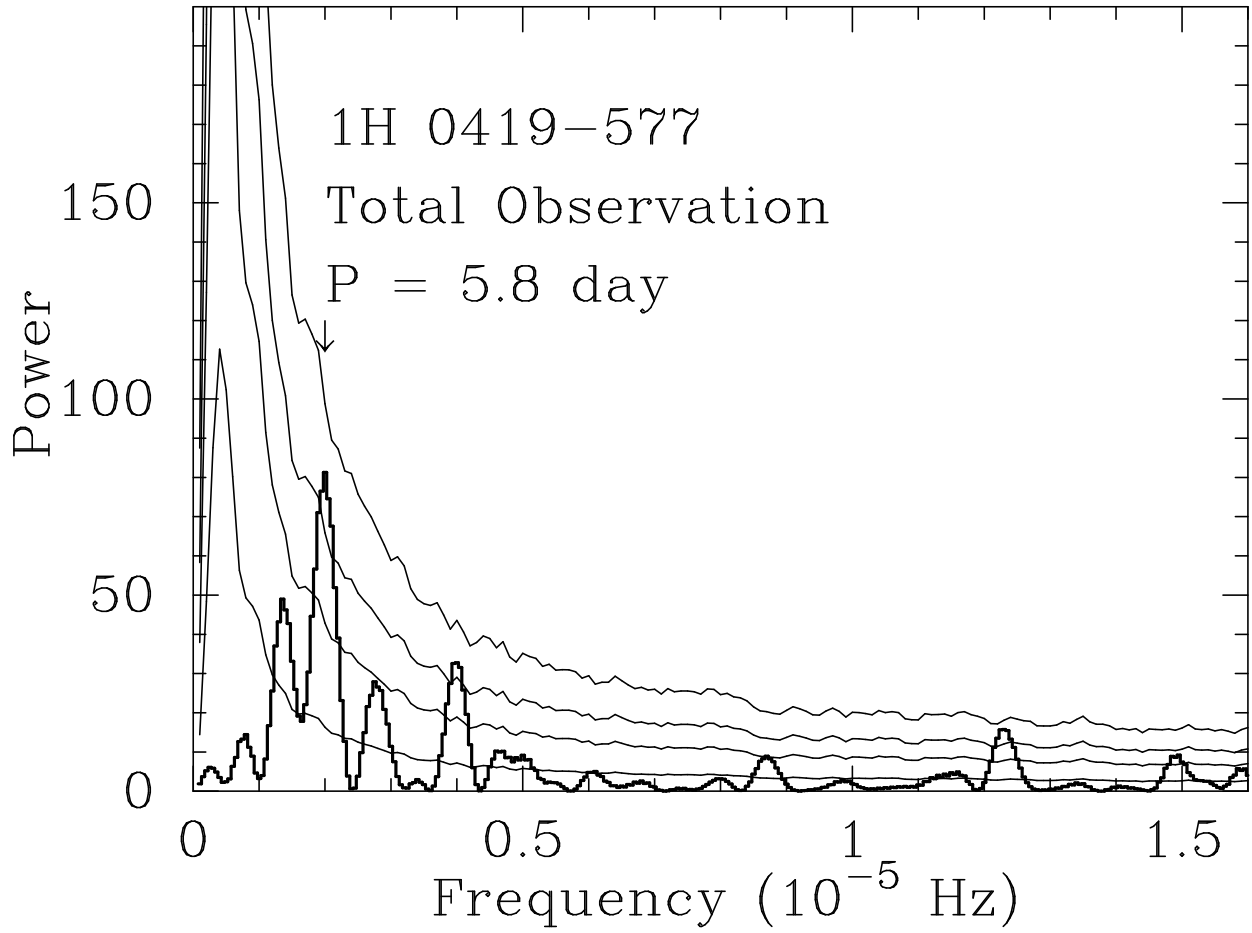}}}
\caption{Periodogram of the \euve\ DS light curve of
1H 0419--577. Lines have the same meaning as in Figure \ref{ps1}.
\label{ps3}}
\end{figure}

In this section, we attempt to quantify the statistical significance
of the periodic signals seen in the periodograms of \sey, Ton~S180,
and 1H~0419--577 in Figures \ref{ps1}--\ref{ps3}.
Perhaps the closest approximation to a formally correct analysis of
significance is that of \cite{tk95}, as implemented by \cite{bwe01}
\footnote{http://astro.uni-tuebingen.de/software/idl/aitlib/}.
\cite{tk95} correctly take into account randomness in 
both Fourier amplitudes and phases in simulating the
power spectra of flickering [$P(F) \propto f^{-1}$]
or random-walk [$P(F) \propto f^{-2}$] light curves.
These processes are good descriptions of many
AGN power spectra which, in the range of frequencies sampled here,
typically follow the form
$P(F) \propto f^{-\alpha}$ where $1 < \alpha < 2$.
We apply this method to estimate the statistical
significance of tentative periodic signals seen in the periodograms
of \sey, Ton~S180, and 1H~0419--577, which are the three
Seyferts that \euve\ observed for at least 20 days each in continuous
or nearly continuous stretches.
The approach involves simulating many red-noise light curves that match
the real data in sampling time, count rate, and variance, and comparing
their periodograms to the periodograms of the real data.  
For a discrete peak in an observed periodogram to be
significant, it must have greater power than nearly all of the simulations at 
that frequency.

In order to determine what red-noise power-law index $\alpha$ matches a 
particular data set, we followed a procedure similar to that used by \cite{l99a}
and \cite{do92}, and essentially the same as \cite{ump02}.    
We simulated 100 light curves for each
of a range of values of $\alpha$ and standard deviation 
$\sigma$ of the light curve that are related to the integrated power spectrum of the data.
Each simulated light curve was constructed as follows.
We first generated a time series with 10 s time resolution that is longer by
up to a factor of 2 than an actual \euve\ light curve. 
For example, to simulate light curves for Ton~S180 a time series
of 524,288 points 
was required.  A light curve was then drawn with random starting time 
from this time series, then resampled and binned exactly as the \euve\ light 
curve.  It was then rescaled by adding the average count rate so
that Poisson noise could be added by replacing each point by one drawn
from a Poisson distribution having the same number of counts. 
Periodograms for each of these 100 light curves were computed.  
These periodograms were rebinned logarithmically, and the
average and standard deviation at each frequency were computed. They
were then compared using $\chi^2$ with the similarly-rebinned power
spectrum from the observed light curve.
The minimum in $\chi^2$ was found as a function
of $\alpha$ and total variance.  Minimum $\chi^2$ were obtained for
$\alpha = 1.55$, 1.3, and 1.3 for \sey, Ton~S180, and 1H~0419--577,
respectively.  We note that the minima are generally fairly broad,
so the range in $\alpha$ and $\sigma$ must contribute to 
uncertainty in the final significance of the candidate signals.

Finally, 1,000 simulated light curves were generated for each object using
its best-fitted values of $\alpha$ and $\sigma$ of the light curve,
their periodograms were generated, and the resulting exponential
probability distribution of power at each frequency was calculated.
Upper limits on the simulated power at various confidence levels are graphed
in comparison with the actual periodograms in Figures \ref{ps1}--\ref{ps3}.
For example,
the 99.9\% confidence limit for Ton~S180 is below the observed peak at 2.08 days,
which means that fewer than 0.1\% of the simulations had as much power at 2.08 days as is observed.  
In Figure \ref{ps2}, one can also see features
in the simulations of Ton S180
near a period of 1 day and its harmonic that are caused by gaps in
light curve where bad points in the observation were removed; this process
has no effect on the significance of the detected signal at 2.08 days.
We note that the way simulated periodograms are created 
for this application constitutes
a conservative test in that we distribute {\it all\/} of the observed
variance in the light curve among the red-noise components, whereas some of
the observed variance could be due to the periodic signal that we are
evaluating, if it is real.

The formal, single-trial probability for chance occurrence of the 2.08 day
signal in Ton~S180
is $1.6 \times 10^{-4}$.  However, one must also take into account the number
of independent frequencies searched and the possibility that a false signal
might arise at any one of these.  In the analysis of Ton S180, we searched
128 independent frequencies from $3.5 \times 10^{-7}$ s$^{-1}$
to $4.6 \times 10^{-5}$ s$^{-1}$, which is half the Nyquist frequency for
a sampling interval of one satellite orbit.  While there is no rigorous
way of extending the single-trial significance to other frequencies that
have different mean power levels, we can approximate the ``global
significance'' of the detection in a manner similar to \cite{bwe01},
by multiplying the single-trial chance probability by the number of
frequencies searched.  In this case, the global significance becomes 98\%.
Similar analysis shows that the single-trial period detection is significant 
at $99.9\%$ confidence for \sey\ , but only in the second half of
the observation (Figure \ref{ps1}b).
If the entire light curve is used, the
significance of its periodic signal drops to $99.7\%$ (Figure \ref{ps1}a);
it is clearly coming from just the second half of the observation.
The global significance corresponding to Figure \ref{ps1}b is 96\%.
Period detection in 1H~0419--577 is significant at $99.6\%$ for a single trial,
but only 64\% globally.  Thus, we find that two of the three candidate periods
have global significance $> 95\%$, a tantalizing if not entirely conclusive result.

\section {Theoretical Implications}

One notable pattern among the candidate periods detected here in
three objects is their ordering as a function of
redshift and luminosity.  The flux of Ton~S180 is approximately
twice that of \sey, and its redshift is slightly higher.  Thus,
the luminosity of Ton~S180 is about 2.5 times that of \sey.
If the luminosity of the source is proportional to the mass of the
black hole, which in turn is proportional to the characteristic (dynamical)
time scale, then
one might expect the period of Ton~S180 (2.08 days) to be about 2.5 times that
of \sey\ (0.89 days), not far from the ``observed'' factor of 2.3.
Also, since the fluxes detected by \euve\
from 1H 0419--577 and \sey\  are approximately the same, their
luminosities scale as $z^2$.  Then
one might expect the period of 1H 0419--577 (5.8 days) to be about 4 times that
of \sey, not far from the ``observed'' factor of 6.  It is encouraging
that these period values appear to be related to the luminosities of the objects,
a physical quantity, and not, for example, to the length of the observation,
which might have favored a more prosaic
explanation in terms of the frequently deceptive
properties of red noise.

A striking property of these
\euve\ light curves is the large amplitude and rapid variability of some
of them.  Although it is generally true
that variability amplitude in AGNs
increases with increasing photon energy, it is
now clear that EUV variability is as dramatic as any detected
at higher energies.  Therefore, it is of fundamental importance
to measure variability in the EUV
because of the likelihood that this component contains the bulk
of the emission from the inner accretion disk, and most of the
bolometric luminosity as well.  This is certainly true in the case of \sey.
Because of its steep power-law spectrum of $\Gamma = 2.2-2.6$
as measured by \ro\ and \asca\ \citep{hm96,woc98},
the soft X-ray variability of \sey\ cannot be attributed to reprocessing
of harder X-rays.  The hard X-ray flux is less than that
of the variable soft X-ray component.  The same is true of two
additional steep spectrum, highly variable objects, Ton~S180 and Mrk~478
\citep{l99a,mcs96,tu02}.  Thus, we are probably viewing with \euve\
the intrinsic variability of the innermost part of the accretion disk,
the primary energy source.
As hypothesized by \cite{gml93} for the {\it EXOSAT\/} sample,
the absence of a substantial electron-scattering corona,
which would otherwise smooth out the intrinsic variations,
may be what allows us to see large amplitude variability in the 
soft-spectrum EUV sources.  In this picture,
those objects with flat X-ray spectra are the ones that possess
the Comptonizing layer which is needed to produce the hard X-ray reflection
component, and which also diminishes variability if present.

Theory is in fact quite challenged to produce the large-amplitude EUV variability
that we observe.  Because of the energetics argument
mentioned above, one cannot beg the question of the mechanism of variability,
as is often done, by appealing to reprocessing of harder X-rays.
``The buck stops here.'' The EUV is where we must search for the
the fundamental reason why AGNs vary, and it would
be helpful if we had a characteristic time scale and amplitude to work with.
For example, the most promising 
``diskoseismic'' theory \citep{nw93,n97} predicts
that radial {\it g}-modes will be trapped at the inner edge of a relativistic
accretion disk, and that the observationally relevant frequency is
$f = 714\ (M_{\odot}/M)\ F(a)$~Hz
where $F(a)$ ranges from 1 to 3.44 as the dimensionless black-hole
angular momentum parameter $a$ ranges from 0 to 0.998.  For a
$10^8\,M_{\odot}$ black hole, the predicted period of oscillation
ranges from 1.62 days for a Schwarzschild black hole to
0.47 days for a maximal Kerr black hole.
This theory of accretion disk oscillations, which has
been applied to Galactic black-hole binaries and AGNs \citep{nl98,p97}, is capable of generating
the appropriate {\it periods} of a few hours to days in AGNs,
but it cannot account
for {\it amplitudes} of more than a few percent.  Thus, while it
even has some trouble
in accounting for the variability of Galactic microquasars, the theory of
accretion disk oscillations is further strained to try to
explain the large X-ray variability amplitude observed in Seyfert galaxies.

Like Ton~S180 and Mrk~478 above,
some of the most dramatically variable X-ray sources
are NLS1s.  These are now well established as
having steeper X-ray spectra than ordinary, broad-line
Seyfert nuclei.  Thus, they are ubiquitous in soft X-ray selected samples
such as those of \ro.
In \ro\ studies of selected NLS1s, \cite{bh97} and \cite{bbf99}
hypothesized that
relativistic beaming by orbiting asymmetries in the inner accretion disk
must be partly responsible for their variability, which sometimes
implies an efficiency greater than that possible for an
isotropically emitting accreting source.  If so, there should be at
least some evidence for periodicity in their light curves if the
emitting structures retain their coherence for several orbits.
It is that sort of quasi-periodic signal that we may be seeing
in these long \euve\ light curves.  However, we are still far from
developing a physical theory for what creates such structures.
It appears that theory will continue to lag behind observation
in this field until and unless some more detailed
pattern to the observations emerges.

\section{Conclusions}

The foundation of the supermassive black hole model for AGNs rests 
squarely on their rapid X-ray variability, which establishes the 
compact nature of these luminous objects beyond a reasonable doubt.
Although AGN variability has been {\it exploited\/} for various purposes,
we have barely begun to address {\it how} and {\it why} AGNs vary. 
The numerous ``reverberation mapping'' campaigns have not,
as a by-product, shed much light
on this question.  Simultaneous, multiwavelength monitoring
has cast a pall over the whole enterprise, because it has
become apparent that there is no simple relation among the
light curves in the various bands, e.g., the studies of NGC 7469
\citep{n98} and NGC 3516 \citep{e00}.  In NGC 5548, there is some evidence that
the underlying variability process is displayed in the EUV
\citep{crb00}.  This might be true generally for Seyfert nuclei
in which the peak of the spectral energy distribution is in the EUV,
but we still don't know how to interpret that variability.  
A fundamental obstacle to understanding is the apparent
lack of a characteristic time scale, a period or quasi-period
that could be interpreted in terms of a dynamical or other
effect related to the size of the putative
inner accretion disk and the mass of the black hole.
It is difficult to interpret the aperiodic variability
that is characteristic of all accretion-powered objects
including AGNs.  But if X-ray periods or quasi-periods could be found in just
a few AGNs, they would provide a quantitative reference point for theory,
just as QPOs in Galactic X-ray transients do for stellar-mass black holes.

The lack of any confirmed X-ray periods in AGNs might be taken as
evidence that they don't exist.  On the other hand, the tentative
detections reported here suggest that we may not yet have observed sufficiently
on the required time scales.  All three \euve\ observations that
were 20 days or more in duration show some indication of periodic behavior.
Evidence is mounting that a dedicated search for
periodic phenomena in Seyfert galaxies could be productive.  However,
unless this goal is considered worthly of another small, specialized mission,
it is unlikely that the door that \euve\ left ajar to new
discoveries in this field will be re-entered in the near future.

\acknowledgments

This work was supported by NASA ADP grant NAG 5--9094 to JPH.
Much of the data presented in this paper were obtained from the Multimission Archive
at the Space Telescope Science Institute (MAST). STScI is operated by 
the Association of Universities for Research in Astronomy, Inc., under NASA 
contract NAS5-26555. Support for MAST for non-HST data is provided by the NASA 
Office of Space Science via grant NAG5-7584 and by other grants and contracts.
We thank the entire staff of the Center for Extreme-Ultraviolet Astrophysics,
whose dedicated operation of the \euve\ satellite and careful processing
of the resulting data motivated and enabled us to produce this Atlas.
We also thank the referee for insightful comments, which led to significant
improvements in this paper.

\clearpage

\clearpage

\clearpage

\clearpage

\clearpage

\clearpage

\clearpage

\clearpage

\clearpage 

\clearpage

\begin{deluxetable}{lcclccc}
\tabletypesize{\scriptsize}
\tablecaption{Log of Observations \label{tbl1}}
\tablewidth{0pt}
\tablehead{\colhead{Object} & $z$ & \colhead{Type} &
\colhead{Dates} & \colhead{MJD} & \colhead{Figure} & \colhead{References}}
\startdata
NGC 4051         & 0.002 & NLS1  & 1996 May 20--29            & 50223--50232 & 1a & 1,2  \\
NGC 4051         & 0.002 & NLS1  & 1996 Dec 11--14            & 50428--50431 & 1b & 1,2  \\
NGC 4051         & 0.002 & NLS1  & 1998 May 8--15             & 50941--50948 & 1c & 3    \\
NGC 4051         & 0.002 & NLS1  & 2000 March 23--28          & 51626--51631 & 1d &      \\
NGC 4151         & 0.003 & Sy1.5 & 1997 Apr 30 -- May 7       & 50568--50575 & 1e &      \\
NGC 5548         & 0.017 & Sy1   & 1993 Mar 10--24            & 49056--49070 & 2a & 4,5  \\
NGC 5548         & 0.017 & Sy1   & 1993 Apr 26 -- May 4       & 49103--49111 & 2b & 4,5  \\
NGC 5548         & 0.017 & Sy1   & 1996 Jun 26 -- Jul 7       & 50260--50271 & 2c &      \\
NGC 5548         & 0.017 & Sy1   & 1998 Jun 18--23            & 50982--50987 & 2d & 6    \\
Mrk 279          & 0.030 & Sy1   & 1994 Apr 22 -- Apr 29      & 49464--49471 & 2e & 7    \\
RE J1034+396     & 0.042 & NLS1  & 1994 Apr 14--20            & 50552--50558 & 3a & 8    \\
RX J0437.4--4711 & 0.053 & Sy1   & 1994 Jan 31 -- Feb 4       & 49383--49387 & 3b &      \\
RX J0437.4--4711 & 0.053 & Sy1   & 1994 Oct 23 -- Nov 12      & 49648--49668 & 3c & 9    \\
RX J0437.1--4731 & 0.144 & NLS1  & 1994 Oct 23 -- Nov 12      & 49648--49668 & 3d &      \\
RX J0436.3--4714 & 0.361 & NLS1  & 1994 Oct 23 -- Nov 12      & 49648--49668 & 3e &      \\
Ton~S180         & 0.062 & NLS1  & 1995 Jul 18--24            & 49916--49922 & 4a & 7    \\
Ton~S180         & 0.062 & NLS1  & 1996 Jul 8--17             & 50271--50281 & 4b &      \\
Ton~S180         & 0.062 & NLS1  & 1999 Nov 12 -- Dec 15      & 51494--51527 & 4c & 10,11 \\
Mrk 478          & 0.079 & NLS1  & 1993 Apr 9--18             & 49086--49095 & 4d & 7,12 \\
Mrk 478          & 0.079 & NLS1  & 1995 Jul 2--5              & 49900--49903 & 4e &      \\
1H 0419--577     & 0.104 & Sy1   & 1997 Dec 15 -- 1998 Jan 10 & 50797--50823 & 5a & 13   \\
3C 273           & 0.158 & QSO   & 1994 Jan 8--14             & 49360--49366 & 5b & 14   \\
3C 273           & 0.158 & QSO   & 1994 Jun 11--18            & 49514--49521 & 5c &      \\
3C 273           & 0.158 & QSO   & 1995 Jan 3--15             & 49720--49732 & 5d & 14   \\
3C 345           & 0.593 & QSO   & 1999 Jun 20--27            & 51349--51356 & 5e &      \\
E 1332+375       & 0.438 & QSO   & 1996 Feb 12--21            & 50125--50134 & 6a & 15   \\
\enddata

{\apj References.}---(1)Uttley et al. 2000;(2)Cagnoni et al. 1999;(3)Uttley et al. 1999;(4)Kaastra et al. 1995;

(5)Marshall et al. 1997;(6)Chiang et al. 2000;(7)Hwang \& Bowyer 1997;(8)Puchnarewicz et al. 2001;

(9)Halpern \& Marshall 1996;(10)Turner et al. 2002;(11)Edelson et al. 2002;(12)Marshall el al. 1996;

(13)Halpern et al. 1998;(14)Ramos et al. 1997;(15)Christian et al. 1999.

\end{deluxetable}

\clearpage 

\end{document}